\documentclass[twocolumn, prl, amsmath,amssymb, aps]{revtex4-1}

\usepackage{graphicx}
\usepackage{amsmath,amssymb}
\providecommand{\ang}[1]{\langle#1\rangle}

\begin{document}

%\preprint{}
 
\title{Phase diagram for the Kuramoto model with van Hemmen interactions}


\author{Isabel M. Kloumann, Ian M. Lizarraga, and Steven H. Strogatz}
%\thanks{}
\affiliation{Center for Applied Mathematics, Cornell University, Ithaca, New York 14853}


\date{\today}

\begin{abstract}
We consider a Kuramoto model of coupled oscillators that includes quenched random interactions of the type used by van Hemmen in his model of spin glasses. The phase diagram is obtained analytically for the case of zero noise and a Lorentzian distribution of the oscillators' natural frequencies. Depending on the size of the attractive and random coupling terms, the system displays four states: complete incoherence, partial synchronization, partial antiphase synchronization, and a mix of antiphase and ordinary synchronization. 
\end{abstract}


\maketitle

%\section{Introduction}
%\label{sec:intro}

In 1967, Winfree~\cite{Winfree1967} discovered that synchronization in large systems of coupled oscillators occurs cooperatively, in a manner strikingly analogous to a phase transition. In this analogy, the temporal alignment of oscillator phases plays the same role as the spatial alignment of spins in a ferromagnet. Since then, Kuramoto and many other theorists have deepened and extended this analogy~\cite{kuramoto1984chemical,RevModPhysKuramoto,Strogatz20001,pikovsky2003synchronization,strogatz2003sync}. 

Yet one question has remained murky.  Can a population of oscillators with a random mix of attractive and repulsive couplings undergo a transition to an ``oscillator glass'' \cite{Daido1987}, the temporal analog of a spin glass \cite{binderyoung1986}? Daido \cite{Daido1992} simulated an oscillator analog of the Sherrington-Kirkpatrick spin-glass model \cite{sherrington1975} and reported evidence for algebraic relaxation to a glassy form of synchronization \cite{stillerradons1998,daido2000,stillerradons2000}, but those results are not  yet understood analytically. Others have looked for oscillator glass in simpler models with site disorder (where the randomness is intrinsic to the oscillators themselves, not to the couplings between them) \cite{Daido1987,bonilla1993glassy,Paissan2008,iatsenko2013oscillatorGlass,strogatzhongprl}. Even in this setting the existence of an oscillator glass state remains an open problem. 

In this paper we revisit one of the earliest models proposed for oscillator glass \cite{bonilla1993glassy}: a Kuramoto model whose attractive coupling is modified to include quenched random interactions of the form used by van Hemmen in his model of spin glasses \cite{vanhemmen1982spinGlass}.  The model can now be solved exactly, thanks to a remarkable ansatz recently discovered by Ott and Antonsen \cite{ottAntonsen2008ansatz}. Their breakthrough has already cleared up many other longstanding problems about the Kuramoto model and its offshoots \cite{childs2008,pikovskyrosenblum2008,ottanstonsen2009,martens2009,laing2009,strogatzhongprl,komarov2011,montbriopazo2011,iatsenko2013ensemble,iatsenko2013oscillatorGlass}. For the Kuramoto-van Hemmen model examined here, the Ott-Antonsen ansatz reveals that the model's long-term macroscopic dynamics are reducible to an eight-dimensional system of ordinary differential equations.  Two physically important consequences are that the model does not exhibit algebraic relaxation to any of its attractors, nor does it have the vast number of metastable states one would expect of a glass. On the other hand, the frustration in the system does give rise to two states whose glass order parameter is non-zero above a critical value of the van Hemmen coupling strength. Our main results are exact solutions for the model's macroscopic states, their associated order parameters, and the phase boundaries between them.

The governing equations of the model are
\begin{eqnarray}
\dot\theta_i =\ & \omega_i + \sum_{j = 1}^N K_{ij} \sin(\theta_j - \theta_i) 
\label{eqn:model}
\end{eqnarray}
for $i = 1, \dots, N \gg 1$, where
\begin{eqnarray}
K_{ij} = \dfrac{K_0}{N} + \dfrac{K_1}{N} (\xi_i \eta_j + \xi_j \eta_i).
\label{eqn:coupling}
\end{eqnarray}
Here $\theta_i$ is the phase of oscillator $i$ and $\omega_i$ is its natural
frequency, randomly chosen from a Lorentzian distribution
of width $\gamma$ and zero mean: $g(\omega) = \gamma/[\pi (\omega^2 +
\gamma^2)]$. By rescaling time, we may set $\gamma = 1$ without loss of
generality. The parameters $K_0,\ K_1 \geq 0$ are the Kuramoto and van Hemmen
coupling strengths, respectively. The random variables $\xi_i$ and $\eta_i$
are independent and take the values $\pm 1$ with equal probability.  

Simulations of the model (Fig.~\ref{finiteNstates}) show four types of long-term behavior. (1) \emph {Incoherence} (Fig.~\ref{finiteNstates}(a)): When $K_0$ and $K_1$ are small, the oscillators run at their natural frequencies and their phases scatter. (2) \emph {Partial locking} (Fig.~\ref{finiteNstates}(b)): If we increase $K_0$ while keeping $K_1$ small, oscillators in the middle of the frequency distribution lock their phases while those in the tails remain desynchronized.  (3) \emph {Partial antiphase locking} (Fig.~\ref{finiteNstates}(c)): If instead we increase $K_1$ while keeping $K_0$ small, the system settles into a state of partial antiphase synchronization, where half of the central oscillators lock their phases 180 degrees apart while the other half behaves incoherently. (4) \emph {Mixed state} (Fig.~\ref{finiteNstates}(d)): If both $K_0$ and $K_1$ are sufficiently large and in the right proportion, we find a mixed state that combines aspects of the partially locked and antiphase locked states. But note two changes---the central oscillators that behaved incoherently in Fig.~\ref{finiteNstates}(c) now lock as in Fig.~\ref{finiteNstates}(b), and the antiphase locked oscillators of  Fig.~\ref{finiteNstates}(c) are now less than 180 degrees apart.

%%%%%%%%%%%  Fig 1 %%%%%%%%%%%%
\def\ksmall{1}
\def\klarge{2.5}
\def\klarger{2.75}

\begin{figure}[htbp!]
\includegraphics{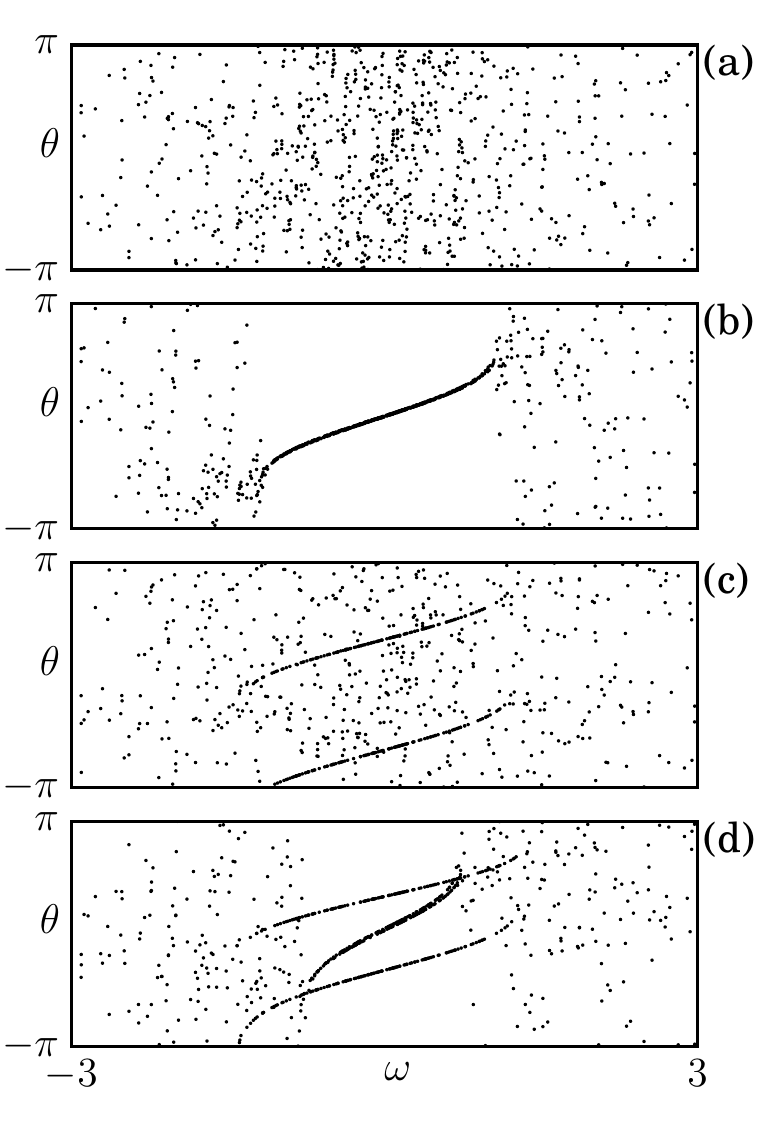}
\caption{\label{finiteNstates} 
Statistical steady states for the Kuramoto-van Hemmen model.  Equation~\eqref{eqn:model}
was integrated numerically for  $N = 1000$ oscillators with Lorentzian distributed frequencies and random initial
phases, using a fourth-order Runge-Kutta method with a fixed step size of 0.05.
Parameter values: (a) Incoherence: $K_0=\ksmall, K_1=\ksmall$; (b) Partial locking: $K_0=\klarge,
K_1=\ksmall$; (c) Partial antiphase locking: $K_0=\ksmall, K_1=\klarger$; (d) Mixed state: $K_0=\klarge, K_1=\klarger$. Only oscillators with $ -3 \leq \omega \leq 3$ are shown. }

\end{figure}
%%%%%%%%%%%%%%%%%%%%%%%%%%

\begin{figure}[htpb!]
\includegraphics{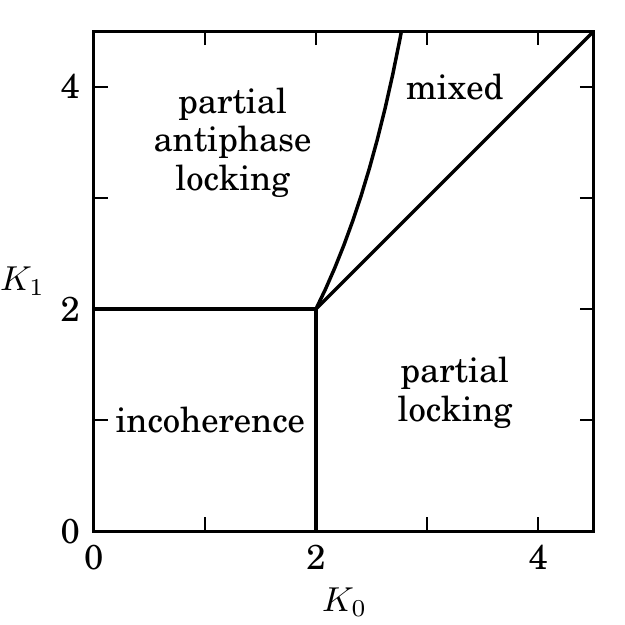}
\caption{Phase diagram for \eqref{eqn:model}, \eqref{eqn:coupling} with $g(\omega) = 1/[\pi (1+ \omega^2)]$.} 
{\label{phaseDiagram}} 
\end{figure}

These four states are not new. They were found and analyzed by Bonilla et al. \cite{bonilla1993glassy} for a variant of Eq.~\eqref{eqn:model} with a white noise term and a uniform (not Lorentzian) distribution of natural frequencies. The advantage of the present system is that the stability properties and phase boundaries of the four states  can be obtained analytically. Figure~\ref{phaseDiagram} shows the resulting phase diagram.

%%%%%%%%%%%%%  Fig 2   %%%%%%%%%
%%%%%%%%%%%%%%%%%%%%%%%%%%

We turn now to the analysis. As mentioned above, the Ott-Antonsen ansatz \cite{ottAntonsen2008ansatz} has become standard, so we suppress the intermediate steps in the following derivation (but see \cite{ottAntonsen2008ansatz} for details).
The ansatz applies to \eqref{eqn:model} in the continuum limit and restricts attention to an invariant manifold that determines the system's long-term dynamics \cite{ottanstonsen2009}. On this manifold the time-dependent density $\rho(\theta,t, \omega, \xi, \eta)$ of oscillators at phase $\theta$ with natural frequency $\omega$ and van Hemmen parameters $\xi, \eta$ is given by
\begin{eqnarray}
\rho = \frac{1}{2\pi}\left\{ 1 + \left[\sum_{n=1}^{\infty} (\alpha^* e^{i\theta })^n  + {\rm ~c.c.}\right]\right\} 
\label{eqn:OAansatz}
\end{eqnarray}
where $\alpha = \alpha(t,\omega, \xi, \eta)$ and the asterisk and c.c. denote complex conjugation. This density evolves according to
\begin{eqnarray}
\frac{\partial \rho}{\partial t} + \frac{\partial}{\partial \theta}(\rho v) &=& 0\label{eqn:conteqn}
\end{eqnarray}
where $v = v(t,\omega, \xi, \eta)$ denotes the velocity field in the continuum limit,
\begin{eqnarray}
v = \omega + {\rm Im}[e^{-i \theta} (K_0 Z + K_1 \xi W_{\eta} + K_1 \eta W_{\xi}) + {\rm ~c.c.}] \label{eqn:vfield}
\end{eqnarray}
and the complex order parameters $Z$, $W_{\xi}$, and $W_{\eta}$ are
\begin{eqnarray}
Z &=& \ang{e^{i\theta}},\nonumber\\
W_{\xi} &=& \ang{\xi e^{i \theta}}, \nonumber\\
W_{\eta} &=&  \ang{\eta e^{i \theta}}.\label{eqn:ordparams}
\end{eqnarray}

The angle brackets $\ang{\cdot}$ denote integration with respect to the
probability measure $\rho(\theta) d \theta \,g(\omega) d\omega \,p(\xi) d\xi
\,p(\eta) d\eta $. The distribution $p$ is normalized so that
$\xi$ and $\eta$ equal $\pm 1$ with equal probability $\frac{1}{2}$.
 
When \eqref{eqn:OAansatz} and \eqref{eqn:vfield} are inserted into
\eqref{eqn:conteqn}, one finds that the dependence on $\theta$ is satisfied
identically if $\alpha(t,\omega, \xi, \eta)$ evolves according to:
\begin{eqnarray} \dot{\alpha} &=& -\frac{\alpha^2}{2}\left[K_0 Z^* + K_1
\left(\xi W_{\eta}^* + \eta W_{\xi}^*\right)\right] + i\omega \alpha
\nonumber\\ &&+ \frac{1}{2}[K_0 Z + K_1 (\xi W_{\eta} + \eta
W_{\xi})].\label{eqn:alphadot} \end{eqnarray} This system is
infinite-dimensional, since there is one equation for each real $\omega$. But
its macroscopic dynamics are governed by a much smaller, finite-dimensional
set of ODEs. The reduction occurs because the different $\alpha(t,\omega, \xi,
\eta)$ in  \eqref{eqn:alphadot} are coupled only through the order parameters
$Z$, $W_{\xi}$, and $W_{\eta}$. Those order parameters in turn are
expressible, via \eqref{eqn:ordparams}, as integrals involving $\rho$ and
therefore $\alpha$ itself.   Under the usual analyticity assumptions
\cite{ottAntonsen2008ansatz} on $\alpha$, the various integrals can be
expressed in terms of a finite set of $\alpha$'s, and these obey the promised
ODEs, as follows.

Consider $Z = \int e^{i\theta} \rho(\theta) d \theta \,g(\omega) d\omega \,p(\xi) d\xi \,p(\eta) d\eta $. To calculate this multiple integral, first substitute \eqref{eqn:OAansatz} for $\rho$ and perform the integration over $\theta$ to get  $Z = \int \alpha \,g(\omega) d\omega \,p(\xi) d\xi \,p(\eta) d\eta $. Second, evaluate the integral $\int_{-\infty}^{\infty}  \alpha\,g(\omega) d \omega$ by considering $\omega$ as a complex number and computing the resulting contour integral, choosing the contour to be an infinitely large semicircle closed in the upper half plane. The Lorentzian $g(\omega) = 1/[\pi (1+ \omega^2)]$ has a simple pole at $\omega = i$, so the residue theorem yields
\begin{eqnarray}
  \int_{-\infty}^{\infty}\alpha\, g(\omega) d\omega &=& \alpha (t, i, \xi, \eta). \label{eqn:omegaintegral} 
\end{eqnarray}
Third, integrate over $\xi$ and $\eta$. Since these variables take on the values 
$\pm 1$ with equal probability, $Z$ receives contributions from four
subpopulations: $(\xi, \eta)$=$(+1,+1)$, $(+1,-1)$, $(-1,+1)$, and $(-1,-1)$.
If we define the sub-order parameters for these subpopulations as
\begin{eqnarray}
A(t) &=& \alpha(t, i, +1,+1)\nonumber\\
B(t) &=& \alpha(t, i, -1, -1)\nonumber\\
C(t) &=& \alpha(t, i,+1,-1) \nonumber\\
D(t) &=& \alpha(t, i, -1, +1), \label{eqn:abcddefine}
\end{eqnarray}
we find that $Z$ is given by
\begin{eqnarray}
Z &=& \frac{1}{4}(A+B+C+D). \label{eqn:Z}
\end{eqnarray}
Similar calculations show that the glass order parameters can also be expressed in terms of $A,B,C,D$:
\begin{eqnarray}
W_{\xi} &=& \dfrac{1}{4}(A-B+C-D), \nonumber\\
W_{\eta} &=& \dfrac{1}{4}(A-B-C+D). \label{eqn:abcdfields}
\end{eqnarray}

The sub-order parameters $A, B, C, D$ have physical meanings. For example, $A$ can be thought of as a giant oscillator, a proxy for all the microscopic oscillators with $(\xi, \eta) = (+1,+1)$. Likewise, $B, C$ and $D$ represent giant oscillators for the other subpopulations. 

The equations of motion for these giant oscillators are obtained by inserting  \eqref{eqn:Z}, \eqref{eqn:abcdfields} into \eqref{eqn:alphadot} and analytically continuing to $\omega = i$. The result is the following closed system:
\begin{eqnarray}
\dot{A} &=& -\frac{1}{2}A^2[K_0 Z^* + \frac{K_1}{2}(A^* - B^*)] -  A \nonumber\\
&& + \frac{1}{2}[K_0 Z + \frac{K_1}{2}(A-B)] \nonumber \\
\dot{B} &=& -\frac{1}{2}B^2[K_0 Z^* + \frac{K_1}{2}(B^* - A^*)] -  B\nonumber\\
&& + \frac{1}{2}[K_0 Z + \frac{K_1}{2}(B-A)] \nonumber \\
\dot{C} &=& -\frac{1}{2}C^2[K_0 Z^* + \frac{K_1}{2}(D^* - C^*)] -  C \nonumber\\
&& + \frac{1}{2}[K_0 Z + \frac{K_1}{2}(D-C)] \nonumber \\
\dot{D} &=& -\frac{1}{2}D^2[K_0 Z^* + \frac{K_1}{2}(C^* - D^*)] -  D \nonumber\\
&& + \frac{1}{2}[K_0 Z + \frac{K_1}{2}(C-D)]. \label{eqn:ABCDsystem}
\end{eqnarray}
Since $A,B,C,$ and $D$ are complex numbers, the system \eqref{eqn:ABCDsystem} is eight-dimensional. 

\begin{figure}[htpb!]
\includegraphics{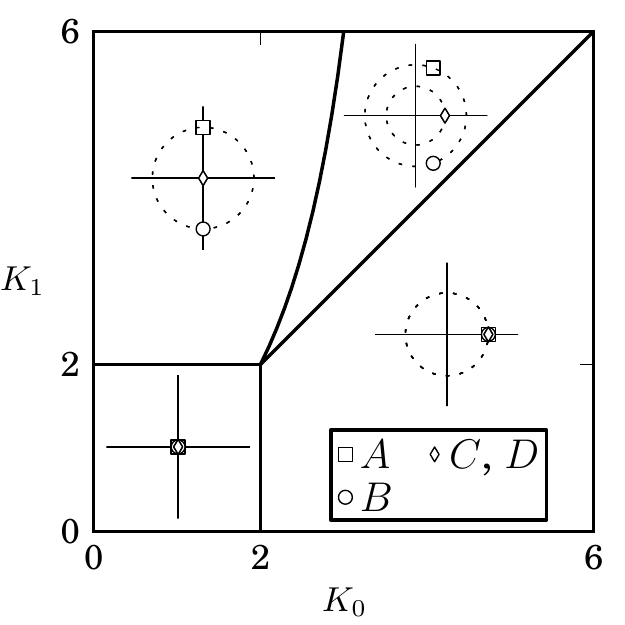}
\caption{Stable fixed points $A, B,C, D$ for the four states. In each panel, the axes show the region of the complex plane with $-1 \leq {\rm ~Re}(z) \leq 1$ and $-1 \leq {\rm ~Im}(z) \leq 1$. Rotationally equivalent fixed points lie on the dashed circles.}
{\label{phaseDiagramWithOrderParameters}} 
\end{figure}

The four steady states shown in Fig.~\ref{finiteNstates} correspond to four families of fixed points of \eqref{eqn:ABCDsystem}, each of which is characterized by a simple configuration of $A, B, C, D$ in the complex plane.  Figure~\ref{phaseDiagramWithOrderParameters} plots those four families schematically on the phase diagram, showing where each exists and is linearly stable. We discuss them in turn. 

%%%%%%%%%%%%% Figure 3 %%%%%%%%%%%%%%
%%%%%%%%%%%%%%%%%%%%%%%%%%%%%%%%

The incoherent state of Fig.~\ref{finiteNstates}(a) corresponds to the fixed point at the origin, $A=B=C=D=0$, with order parameters $Z=W_{\xi}= W_{\eta}=0.$ It exists for all $K_0,\ K_1 \geq 0$ but is linearly stable iff (if and only if) $K_0 < 2$ and $K_1 < 2$. This stability region is shown as the square in the lower left of Fig.~\ref{phaseDiagramWithOrderParameters}. 

The partially locked state (Fig.~\ref{finiteNstates}(b)) corresponds to a
configuration where $A, B, C$ and $D$ all equal the same nonzero complex
number, as shown in the lower right panel of
Fig.~\ref{phaseDiagramWithOrderParameters}. By rotational symmetry, we can assume that $A=B=C=D=R_{PL}>0$. Such a state is a fixed point
of \eqref{eqn:ABCDsystem} iff $K_0 >  2$ and $R_{PL} = \sqrt{1 - 2/K_0}$, in
which case it is linearly stable iff  $K_1 < K_0$. (There is a trivial zero
eigenvalue associated with the rotational symmetry, so what we really mean is
that the state is linearly stable to all perturbations other than rotational
ones. Likewise, there is a whole circle of partially locked states, all
equivalent up to rotation, as indicated by the dashed circle in the lower
right panel of Fig.~\ref{phaseDiagramWithOrderParameters}.) The order
parameters are $Z= \sqrt{1 - 2/K_0}$ and $W_{\xi}= W_{\eta}=0.$

The antiphase state (Fig.~\ref{finiteNstates}(c)) corresponds to a fixed point where $A = -B = R_{A} > 0$  and $C = D = 0$. It exists iff $K_1 > 2$ and $R_{A} =\sqrt{1 - 2/K_1}$. When it exists it is linearly stable iff 
\begin{eqnarray}
\label{antiphase_stability_boundary}
K_0 < 4 K_1/(2 + K_1).
\end{eqnarray}

Finally, the mixed state (Fig.~\ref{finiteNstates}(d)) corresponds to a configuration where $A = B^*$ and $C = D = R_{M} > 0$. It exists iff $K_1 > 2$ and $4 K_1/(2 + K_1) < K_0  < K_1$ (the wedge in the upper right of Fig.~\ref{phaseDiagramWithOrderParameters}) and satisfies 
\begin{eqnarray}
{\rm ~Re}(A) &=& \frac{K_0}{2 K_1 - K_0}\sqrt{1 + \frac{2}{K_1}-\frac{4}{K_0}}\nonumber\\
{\rm ~Im}(A) &=&  2\sqrt{\frac{(K_1 - K_0)(K_1 (K_1 - 2) + K_0 )}{K_1(2 K_1 - K_0)^2}}\nonumber\\
 R_M &=& \sqrt{1 + \frac{2}{K_1} - \frac{4}{K_0}}. \label{eqn:mixedsoln}
\end{eqnarray}
We were unable to find the eigenvalues analytically in this final case, but we verified linear stability numerically for a sample of mixed states up to $K_1 = 10^5$. 

All the transitions in Fig.~\ref{phaseDiagramWithOrderParameters} are
continuous (Fig.~4). In particular, the mixed state morphs
into the antiphase state on the left side of its stability region, and into
the partially locked state on the right side. To verify this, observe that the
configuration of $A,B,C, D$ in the mixed state, as parametrized by
Eq.~\eqref{eqn:mixedsoln}, continuously deforms into the states on either side
of it as $(K_0,K_1)$ approaches the relevant stability boundary.
%%%%%%%%%%%%% Figure 4 %%%%%%%%%%%%%%
\begin{figure}[htpb!]
\includegraphics[width=\linewidth]{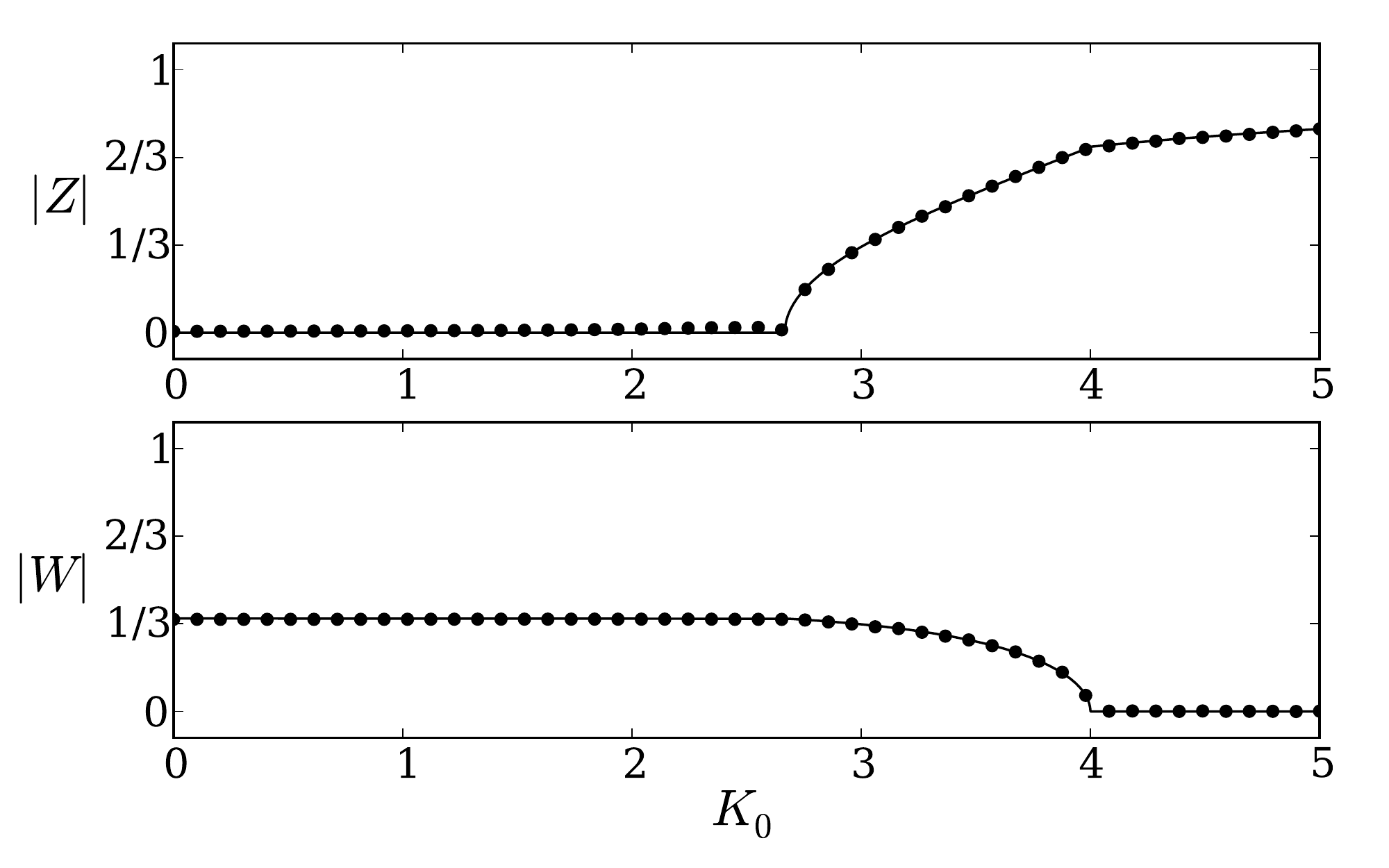}
\caption{Theory vs. simulation for order parameters. Solid line, exact
results; circles, simulations for $N$=$50,000$ oscillators.  
For $K_1$=4, Eq.~\eqref{eqn:model} was integrated using an
Euler method with step size 0.01. Each
combination of $(\xi,\eta) = (\pm 1, \pm 1)$ was assigned $N/4$ oscillators, 
with natural frequencies taken from a deterministic Lorentzian distribution:
$\omega_i$ = $\tan\left[(\pi/2)(2i - n -1)/(n+1)\right]$, 
for $i=1, \dots, n$ and $n$=$N/4$. The values of the order parameters
are shown at $t = 200$, by which time convergence to a statistical steady state
has occurred. }

\label{orderParametersTheorySimulation}
\end{figure}
%%%%%%%%%%%%%%%%%%%%%%%%%%%%%%%%

The glass order parameters $W_\xi$ and $W_\eta$ are nonzero for the antiphase
and mixed states, so in that specific sense the model can be said to exhibit a
glassy form of synchronization~ \cite{bonilla1993glassy}. Moreover, $W_\xi =
W_\eta$ for all four states, which confirms a conjecture of Bonilla et
al.~\cite{bonilla1993glassy}. On the other hand, the oscillator model
\eqref{eqn:model}, \eqref{eqn:coupling} lacks other defining features of a
glass, such as a large multiplicity of metastable states and non-exponential
relaxation dynamics; the same is true of the original van Hemmen spin-glass
model \cite{choySherrington1984vanhemmen}.

Experimental tests of the phase diagram predicted here may be possible in a variety of oscillator systems with programmable coupling. Prime candidates are optical arrays \cite{hagerstrom2012experimental} or populations of photosensitive chemical oscillators \cite{tinsley2012chimera} in which the interactions are mediated by a computer-controlled spatial light modulator.

Research supported in part by an NSF Graduate Research Fellowship to I.M.K and NSF Award 1006272 to I.M.L. We thank Murray Strogatz for helpful discussions. 
%\end{acknowledgments}

\bibliography{refs.bib}

\end{document}